\documentclass{article}
\usepackage{amsmath, amssymb, amsfonts}
\title{ON THE BLACK HOLE INTERIOR SPACETIME}
\author{Hristu Culetu, \\Ovidius University, Dept.of Physics, \\B-dul Mamaia 124, 8700 Constanta, Romania, \\e-mail : hculetu@yahoo.com}

\begin{document}
\numberwithin{equation}{section}
\pagenumbering{arabic}
\maketitle
\newcommand{\fv}{\boldsymbol{f}}
\newcommand{\tv}{\boldsymbol{t}}
\newcommand{\gv}{\boldsymbol{g}}
\newcommand{\OV}{\boldsymbol{O}}
\newcommand{\wv}{\boldsymbol{w}}
\newcommand{\WV}{\boldsymbol{W}}
\newcommand{\NV}{\boldsymbol{N}}
\newcommand{\hv}{\boldsymbol{h}}
\newcommand{\yv}{\boldsymbol{y}}
\newcommand{\RE}{\textrm{Re}}
\newcommand{\IM}{\textrm{Im}}
\newcommand{\rot}{\textrm{rot}}
\newcommand{\dv}{\boldsymbol{d}}
\newcommand{\grad}{\textrm{grad}}
\newcommand{\Tr}{\textrm{Tr}}
\newcommand{\ua}{\uparrow}
\newcommand{\da}{\downarrow}
\newcommand{\ct}{\textrm{const}}
\newcommand{\xv}{\boldsymbol{x}}
\newcommand{\mv}{\boldsymbol{m}}
\newcommand{\rv}{\boldsymbol{r}}
\newcommand{\kv}{\boldsymbol{k}}
\newcommand{\VE}{\boldsymbol{V}}
\newcommand{\sv}{\boldsymbol{s}}
\newcommand{\RV}{\boldsymbol{R}}
\newcommand{\pv}{\boldsymbol{p}}
\newcommand{\PV}{\boldsymbol{P}}
\newcommand{\EV}{\boldsymbol{E}}
\newcommand{\DV}{\boldsymbol{D}}
\newcommand{\BV}{\boldsymbol{B}}
\newcommand{\HV}{\boldsymbol{H}}
\newcommand{\MV}{\boldsymbol{M}}
\newcommand{\be}{\begin{equation}}
\newcommand{\ee}{\end{equation}}
\newcommand{\ba}{\begin{eqnarray}}
\newcommand{\ea}{\end{eqnarray}}
\newcommand{\bq}{\begin{eqnarray*}}
\newcommand{\eq}{\end{eqnarray*}}
\newcommand{\pa}{\partial}
\newcommand{\f}{\frac}
\newcommand{\FV}{\boldsymbol{F}}
\newcommand{\ve}{\boldsymbol{v}}
\newcommand{\AV}{\boldsymbol{A}}
\newcommand{\jv}{\boldsymbol{j}}
\newcommand{\LV}{\boldsymbol{L}}
\newcommand{\SV}{\boldsymbol{S}}
\newcommand{\av}{\boldsymbol{a}}
\newcommand{\qv}{\boldsymbol{q}}
\newcommand{\QV}{\boldsymbol{Q}}
\newcommand{\ev}{\boldsymbol{e}}
\newcommand{\uv}{\boldsymbol{u}}
\newcommand{\KV}{\boldsymbol{K}}
\newcommand{\ro}{\boldsymbol{\rho}}
\newcommand{\si}{\boldsymbol{\sigma}}
\newcommand{\thv}{\boldsymbol{\theta}}
\newcommand{\bv}{\boldsymbol{b}}
\newcommand{\JV}{\boldsymbol{J}}
\newcommand{\nv}{\boldsymbol{n}}
\newcommand{\lv}{\boldsymbol{l}}
\newcommand{\om}{\boldsymbol{\omega}}
\newcommand{\Om}{\boldsymbol{\Omega}}
\newcommand{\Piv}{\boldsymbol{\Pi}}
\newcommand{\UV}{\boldsymbol{U}}
\newcommand{\iv}{\boldsymbol{i}}
\newcommand{\nuv}{\boldsymbol{\nu}}
\newcommand{\muv}{\boldsymbol{\mu}}
\newcommand{\lm}{\boldsymbol{\lambda}}
\newcommand{\Lm}{\boldsymbol{\Lambda}}
\newcommand{\opsi}{\overline{\psi}}
\renewcommand{\tan}{\textrm{tg}}
\renewcommand{\cot}{\textrm{ctg}}
\renewcommand{\sinh}{\textrm{sh}}
\renewcommand{\cosh}{\textrm{ch}}
\renewcommand{\tanh}{\textrm{th}}
\renewcommand{\coth}{\textrm{cth}}

\begin{abstract}
A new version of the geometry inside a black hole is proposed, on the grounds of an idea given by Doran et al. The spacetime is still time dependent and is a solution of Einstein's equations with a stress tensor corresponding to an anisotropic fluid. The model leads to a time-dependent cosmological constant.\\
 The energy density of the fluid is proportional to $1/t^{2}$, as in many dark energy models but the Brown - York quasilocal energy of the black hole equals its mass $m$.

Keywords : anisotropic fluid, time dependent cosmological constant, quasilocal energy, dark energy.
\end{abstract}

\section{INTRODUCTION}

The exterior solution of the Schwarzschild black hole was extensively studied in the last decades. Its fundamental role is related to the phenomenon of gravitational collapse, black hole evaporation, event horizons, spacetime singularities.

Unlike the exterior geometry, the interior of a black hole is hidden behind the event horizon. As Brehme \cite{RB} has remarked since 1977, the fascinating picture of the interior of the black hole, although very strange, is nevertheless understandable and imaginable. The external spatial and temporal coordinates exchange their character when the event horizon is crossed, the interior solution representing a non static spacetime, with time dependent metric coeficients \cite{DLC}. 

Brehme observed that, viewed from inside the horizon, the ''point source'' is rather an ''instant source''. The mass $m$ of the black hole appears only at the moment $t = 0$, vanishing for all later moments. Therefore, the universe for $r < 2m$ ( of Kantowski - Sachs form)
\begin{equation}
ds^{2} = -(\frac{2m}{t}-1)^{-1} dt^{2} + (\frac{2m}{t}-1) dz^{2} + t^{2} d\Omega^{2}
\label{1}
\end{equation}
is established by an initial condition rather than by a boundary condition in space \cite{RB}. $d \Omega^{2}$ stands for the unit 2-sphere line element and z - the spatial coordinate, which no longer represents a radial coordinate inside the black hole \cite{DLC}. In addition, the equation of motion of an object which is initially at rest with respect to the z - coordinate remains at rest, as for an inertial observer in flat space.\\

\section{THE ANISOTROPIC STRESS TENSOR}

 Doran et al. \cite{DLC} studied an interesting case, in the spirit of $gravastars$ models \cite{DV} \cite {GC} \cite {MM} \cite{CFV} for collapsing objects. They considered a time-dependent mass $m(t)$ in eq. (1.1), case in which the metric is no longer a solution of the vacuum Einstein equations. The authors of \cite{DLC} noted that the spacetime acquires an instantaneous Minkowski form for $m(t) = t$, although the curvature is nonvanishing, with a singularity at $t = 0$. 
 
 It is just the case that will be analysed in this paper. 
 
 The line element (1.1) becomes now
 \begin{equation}
 ds^{2} = -dt^{2} + dz^{2} + t^{2} d\Omega^{2} 
 \label{2}
 \end{equation}
 As far as a test particle moves along the z - axis, the geometry is Minkowskian. Otherwise, the metric is time dependent. 
 
 The Einstein tensor $G_{\mu}^{\nu}$ has been computed for an arbitrary time-dependent mass $m(t)$ by Doran et al. They found that 
\begin{equation}
G_{t}^{t} = -\frac{2}{t^{2}} \frac{dm}{dt} = G_{z}^{z},~~~~G_{\theta}^{\theta} = G_{\phi}^{\phi} = -\frac {1}{t} \frac {d^{2}m}{dt^{2}}
\label{3}
\end{equation}
 (we use the conventions $G = c = 1$ and the Ricci tensor is given by $R_{\alpha \beta} = \partial_{\nu} \Gamma_{\alpha \beta}^{\nu}-...$). The spacetime
\begin{equation}
ds^{2} = -\left[\frac{2m(t)}{t}-1\right]^{-1} dt^{2} + \left[\frac{2m(t)}{t}-1\right] dz^{2} + t^{2} d\Omega^{2} 
\label{4}
\end{equation}
is a solution of Einstein's equations if the interior of the black hole is considered to be filled with an anisotropic fluid \cite{MM} \cite{CFV} for which $\rho = -p$, as could be seen from (2.2) ($\rho$ is the energy density and $p$ - the pressure along the z-direction).

Much in the spirit of \cite{FL} \cite{SV}, we consider the spacetime (2.1) is filled with a locally anisotropic fluid with
\begin{equation}
T_{\mu}^{\nu} = \rho \eta_{\mu} \eta^{\nu} + p s_{\mu} s^{\nu} 
\label{5}
\end{equation}
where $\eta_{\mu} = (1, 0, 0, 0)$ is the anisotropic fluid four velocity (the components are in the order $t,z,\theta, \phi$), $s_{\mu}= (0,1,0,0)$ is the unit spacelike vector in the direction of anisotropy \cite{LH}~(with $\eta_{\alpha} \eta^{\alpha} = -1, ~s_{\alpha} s^{\alpha} = 1~ ~and~ s_{\alpha} \eta^{\alpha} = 0$). Hence, we have only one principal stress $p$ and the transverse pressures measured in the ortogonal direction to $s^{\mu}$ are vanishing.

 $T_{\mu}^{\nu}$ from (2.4) may be put on the r.h.s. of Einstein's equations, with $\rho = -p ~(or ~T_{t}^{t} = - T_{z}^{z})$ and $T_{\theta}^{\theta} = T_{\phi}^{\phi} = 0$ (that is in accordance with $G_{\theta}^{\theta} = G_{\phi}^{\phi} = 0$ when $d^{2}m/dt^{2} = 0$, as is the case for the spacetime (2.1)).

Therefore, the particular case $m(t) = t$ shows that the metric (2.1) is a solution of Einstein's equations with
\begin{equation}
\rho = -T_{t}^{t} = \frac{1}{4 \pi t^{2}},~~~~p = T_{z}^{z} =-\frac{1}{4 \pi t^{2}}
\label{6}
\end{equation}

\section{GEODESICS}

 Let us study now the geodesics in the spacetime (2.1). Instead of using the standard equations for geodesics, we start with the Lagrangean 
 \begin{equation}
 \textbf{L} = \frac{1}{2} g_{\alpha \beta} \frac{dx^{\alpha}}{d\lambda} \frac{dx^{\beta}}{d\lambda}
\label{7}
\end{equation}
where $\lambda$ is an affine parameter along the trajectory. \\

a)~\textbf{Timelike geodesics}\\
From the line element (2.1) we have, with $\theta = \pi/2$,
\begin{equation}
\dot{t}^{2} - \dot{z}^{2} -t^{2} \dot{\phi}^{2} = 1
\label{8}
\end{equation}
and from the Euler - Lagrange equations 
\begin{equation}
\frac{\partial\textbf{L}}{\partial\dot{z}} = \dot{z} \equiv p_{z}, ~~~~\frac{\partial\textbf{L}}{\partial{\dot{\phi}}} = t^{2} \dot{\phi} \equiv L
\label{9}
\end{equation}
where an overdot denotes the derivative with respect to the proper time $\tau$ and the conserved quantities $p_{z}$ and $L$ may be interpreted as a momentum along the z-direction per unit mass and the angular momentum per unit mass, respectively. We note that there is no a conserved energy due to the time dependence of the metric (we do not have a timelike Killing vector). Combining (3.2) with (3.3) one obtains
\begin{equation}
z(t) = \frac{\pm p_{z}}{E}\sqrt{t^{2} + a^{2}}
\label{10}
\end{equation}
where $E = \sqrt{1 + p_{z}^{2}}$ plays the role of an energy and $a = L/E$. We have chosen $z(0) = \pm p_{z}L/E^{2}$ as initial condition. We stress that the coordinate $z$ cannot be considered as a radial coordinate inside the black hole (in fact $-\infty < z <\infty$). Moreover, a particle at rest ($p_{z} = 0$) remains always at rest even though the gravitational field is nonvanishing  (the scalar curvature \cite{DV}
\begin{equation}
R_{\alpha}^{\alpha} = \frac{4}{t^{2}} \frac{dm}{dt} + \frac{2}{t} \frac{d^{2} m}{dt^{2}}
\label{11}
\end{equation}
becomes $R_{\alpha}^{\alpha} = 4/t^{2}$ in our case).

For the hyperbolic motion (3.4), the asymptotic velocity is less than unity
\begin{equation}
\frac{dz(t)}{dt}|_{\infty} = \frac{\pm p_{z}}{E}~ <~1. 
\label{12}
\end{equation}
~We see that, for $t >> a$, the function $z(t)$ is linear, like in the flat space. We may write $a = L/E = b p_{z}/E$ where $b = L/p_{z}$ represents the impact parameter of the particle.

Since the test particle has unit mass, $p_{z}$ is its velocity $v$. For $v << 1$, the condition for $a$ to be negligible with respect to $t$ is $t >> bv$. For example, for a test particle with $v = 10^{3} cm/s$ and $b = 100~ cm$, $a \approx 10^{-16} s$. Even for an elementary particle with $v = 10^{8} cm/s$ and $b = 10^{10}~ m$,~ $a \approx 0.1~ s$. Therefore, we reach at the conclusion that the deviation of the trajectory $z(t)$ from the flat space case is efficient only for very short time intervals. That is in accordance with a negligible scalar curvature for long time intervals.

As far as $\phi(t)$ is concerned, we have
\begin{equation}
\frac{d\phi}{dt} = \frac{\pm a}{t \sqrt{t^{2} + a^{2}}}
\label{13}
\end{equation}
The previous equation shows that the time variation of $\phi(t)$ is important only for very short time intervals. Note that, for $t >> a$, $d\phi/dt$ behaves like $a/t^{2}$. On the contrary, $dz/dt$ is $\pm (p_{z}/E)[1-(a^{2}/2t^{2}]$, containing a constant term. Therefore, in the case $t >> a$,the trajectory of the test particle is, practically, along the z - axis, with no angular component. In addition, (3.7) leads to 
\begin{equation}
\phi(t) = \mp \frac{1}{2}~ ln~ \frac{\sqrt{t^{2} + a^{2}}+a}{\sqrt{t^{2} + a^{2}} - a}
\label{14}
\end{equation}
The fact that $\phi(t)$ tends to minus infinity for $t <<a$ is not surprising inside a black hole. Peixoto and Katanaev \cite{PK} showed that the maximally extended BTZ solution coincides with Minkowski spacetime without any singularity and horizon for infinite range of the angle coordinate $\phi$.\\

b)~\textbf{Null geodesics}\\
~ Let us compute now the geodesics $z(t)$ for a massless test particle. From (3.3) and the fact that 
\begin{equation}
\dot{t}^{2} - \dot{z}^{2} - t^{2} \dot{\phi}^{2} = 0
\label{15}
\end{equation}
we obtain, for the projection of the geodesic on the z - axis
\begin{equation}
z(t) = \pm \sqrt{t^{2} + b^{2}}
\label{16}
\end{equation}
with $z(0) = b$ as initial condition. 

The trajectory is again a hyperbola but now the asymptotic velocity is unity, as it should be. It is worth to note that $1/b$ plays the role of the particle (rest-system) acceleration. The null geodesics $\phi(t)$ could be obtained from
\begin{equation}
\frac{d\phi}{dt} = \frac{\pm b}{t\sqrt{t^{2} + b^{2}}}
\label{17}
\end{equation}
with a singularity at $t = 0$. When $b = 0$, $\phi = ~const.$, as it should be. From (3.11) one obtains
\begin{equation}
\phi(t) = \mp \frac{1}{2}~ ln~ \frac{\sqrt{t^{2} + b^{2}} + b}{\sqrt{t^{2} + b^{2}} - b}
\label{18}
\end{equation}

\section{THE COSMOLOGICAL CONSTANT} 
 ~It is interesting to evaluate the time dependent cosmological constant $\Lambda(t)$ according to the prescription given in \cite{HC}, for the spacetime (2.1). 
 
 The Gaussian curvature of the two - surface $\Sigma$, $t = t_{0} , z = z_{0}$, is given by 
 \begin{equation}
 k = \frac{1}{\sigma} R_{\theta \phi \theta \phi}  
 \label{19}
 \end{equation}
 where $\sigma$ is the determinant of the metric
 \begin{equation}
 d\sigma^{2} = t_{0}^{2} (d\theta^{2} + sin^{2}\theta d\phi^{2})
 \label{20}
 \end{equation}
 and $R_{\theta \phi \theta \phi} = t_{0}^{2} sin^{2} \theta$. Hence
 \begin{equation}
 k \equiv \Lambda(t) = \frac{1}{t^{2}}|_{t = t_{0}}
 \label{21}
 \end{equation}
 ~We observe that (4.3) gives the same expression for the cosmological constant as that obtained in \cite{HC} for the Minkowski spacetime and, in a cosmological context, by Bertolami \cite{OB} in a Brans - Dicke theory with a time - dependent scalar field. On the other hand, the previous form of $\Lambda(t)$ is compatible with (2.5), where $\rho (t)$ plays the role of a cosmological constant.
 
 A similar expression for the energy density was obtained recently by Maziashvili \cite{M} who noticed that the existence of time $t$ fluctuating with the amplitude $\delta t$ implies the existence of energy $~(\delta t)^{-1}$ which is distributed uniformly over the volume $t^{3}$. From here, Maziashvili concluded that the background energy density of the universe (the dark energy) should be 
\begin{equation}
\rho \cong \frac{1}{t_{P}^{2} t^{2}}
\label{22}
\end{equation}
which is nothing else but $\rho (t)$ from (2.5) ($t_{P}$ is the Planck time).

Even though the geometry (2.1) is nonstatic, let us compute formally the Brown - York quasilocal energy \cite{LSY} inside the surface  $\Sigma$
\begin{equation}
W = \frac{1}{8 \pi} \int_{\Sigma} K \sqrt{\sigma} d^{2}x
\label{23}
\end{equation}
where K is the trace of the extrinsic curvature of $\Sigma$. A simple calculation gives $K = 2/t$ and, therefore, 
\begin{equation}
W = \frac{1}{8 \pi} \int_{\Sigma} \frac{2}{t_{0}}~ t_{0}^{2}~ sin\theta~ d\theta~ d\phi = t_{0}
\label{24}
\end{equation}
namely, $W = m(t_{0})$. In other words, the energy of the black hole interior, at the time $t_{0}$, equals its mass $m(t_{0})$. With all fundamental constants, it becomes $W = (c^{5}/G)~t_{0}$ (see also \cite{AAP}).

\section{CONCLUSIONS}

A black hole interior spacetime filled with an anisotropic fluid is proposed. The radial pressure is negative but the tangential ones are vanishing. The energy density of the fluid is proportional to $1/t^{2}$ ($t$ - an arbitrary time), as for many dark energy models.
 
 A time dependent cosmological constant is obtained, using a previous prescription in terms of the Gaussian curvature. The time dependent quasilocal energy of the black hole equals its mass $m$.

\end{document}